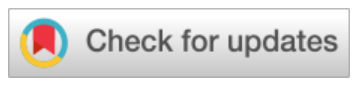

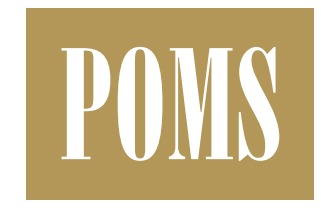

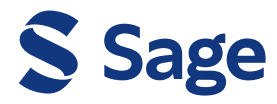

# Merchants of vulnerabilities: How bug bounty programs benefit software vendors

**Esther Gal-Or[1], Muhammad Zia Hydari[1] and Rahul Telang[2]**

## Abstract

We study how bug bounty programs (BBPs) shape software vendors' security and release choices. Vendors invest in internal assurance before release to reduce residual vulnerabilities, and after launch they must manage vulnerability discovery, disclosure, and remediation. We develop a game-theoretic model in which a vendor chooses release timing and severity-contingent bounties, anticipating effort by ethical and malicious hackers in a winner-take-all discovery race. The model highlights two linked mechanisms: an incentive channel that shifts first discovery of severe vulnerabilities away from malicious exploitation and toward ethical reporting, and a governance channel in which coordinated disclosure changes how vulnerability information is managed while remediation is underway. We derive closed-form optimal bounties and characterize a feasibility region that sustains positive bounties and interior success probabilities. Within this region, a BBP strictly increases the vendor's expected profit by reallocating first-discovery probability on severe vulnerabilities from malicious to ethical hackers and by converting part of severe-loss exposure into bounded, pay-for-results expenditures. For private programs, we also solve for the optimal invited set of ethical hackers and show that this optimal set is strictly smaller than the expected number of malicious attackers. Higher bounties raise ethical hackers' effort and first-discovery probabilities but also increase program cost, and they interact with reputational (non-monetary) incentives. Finally, in the baseline model, BBP adoption conditionally reduces the marginal value of additional pre-release delay and therefore conditionally implies earlier release relative to the no-BBP benchmark. This timing result is a within-model conditional implication; its practical relevance depends on operational readiness, triage throughput, and the vendor's ability to validate and safely deploy fixes once a valid report arrives. Managerially, BBPs should be viewed as a post-release governance layer that complements strong internal assurance rather than as a substitute for it. Policymakers can support responsible use of BBPs by encouraging timely remediation, transparent post-patch disclosure, and reporting standards that reduce information asymmetry and triage frictions.

## 1 Introduction

Software vulnerabilities, security flaws exploitable by attackers (Sen et al., 2020), create a recurring tension for software vendors: pressure to shorten time-to-market versus exposure to security losses driven by residual vulnerabilities at release. Vendors mitigate this risk through in-house secure development and testing, as well as post-release processes for receiving, triaging, and remediating vulnerability reports. Bug bounty programs (BBPs) have become a prominent mechanism for structuring post-release vulnerability discovery by incentivizing ethical reporting (Zhang et al., 2025).

We study how BBPs change incentives and information flows in vulnerability discovery, with particular attention to severe vulnerabilities that are subject to a race between ethical reporting and malicious exploitation. We characterize the optimal policy of a profit-maximizing vendor in a stylized environment and use the resulting comparative statics to

[1]Graduate School of Business, University of Pittsburgh, Pittsburgh, PA, USA
[2]Heinz College, Carnegie Mellon University, Pittsburgh, PA, USA

**Corresponding authors:**
Rahul Telang, Heinz College, Carnegie Mellon University, Pittsburgh, PA 15213, USA.
Email: rtelang@andrew.cmu.edu

Muhammad Zia Hydari, Graduate School of Business, University of Pittsburgh.
Email: hydari@alum.mit.edu

explain adoption incentives, severity-contingent bounty pricing, and the marginal effects of BBPs on release timing. The goal of the article is to clarify incentives and market consequences that may not be visible in vendor-authored disclosures, rather than to offer a welfare-optimal policy prescription.

A central managerial puzzle motivating this study is that BBPs simultaneously invite external hackers to probe systems and typically rely on *coordinated disclosure* (also called controlled disclosure): participants report privately to the vendor, and public disclosure is staged or delayed until remediation. Coordinated disclosure can reduce exploit risk while a patch is developed, but it can also increase information asymmetry if users remain uninformed while exposure persists. In the model, BBPs therefore operate through two linked channels: an *incentive* channel that shifts first discovery of severe vulnerabilities from malicious exploitation toward ethical reporting, and a *governance* channel that reduces the vendor's expected cost from uncoordinated public disclosure by conditioning rewards on private reporting. The strength of the governance channel depends on the length of the remediation window and on the vendor's ability to process reports. AI can compress patch-drafting time while simultaneously increasing verification and triage burden, so the net effect of automation on this channel depends on operational readiness and deployment governance. In the limiting case of truly near-instant remediation, the private-information window created by coordinated disclosure becomes short, so the governance role of the BBP correspondingly shrinks even though the discovery-allocation role remains. When triage is congested or deployment is slow, coordinated disclosure matters more for risk containment, but it also prolongs the period of user-facing information asymmetry. This disclosure-governance tension is increasingly salient in practice and motivates our focus on BBPs as arrangements that jointly shape discovery incentives and the timing of information release (Telang and Hydari, 2025).[1]

Existing work studies vulnerability disclosure and BBPs as incentive mechanisms and marketplaces (Feng et al., 2024; Zhang et al., 2025), but there is limited analytical guidance on the joint implications of three features that often co-occur in practice: severity-specific discovery races between ethical and malicious actors, BBP-induced control of disclosure timing and the associated information asymmetry, and release timing as a strategic decision that trades time-to-market against residual vulnerability risk. Our model integrates these elements and yields tractable implications that connect program design to both security outcomes and vendor incentives. Our analysis yields four main results. First, we derive vendor-optimal, severity-contingent bounty levels and show how they vary with the vendor's loss exposure from severe vulnerabilities, the expected gains to malicious exploitation, and the reputational benefits associated with ethical reporting. Second, we characterize conditions under which a BBP is feasible and profitable, and we show how BBPs can shift the probability that severe vulnerabilities are first found by ethical hackers rather than malicious hackers. Third, for private programs we characterize the optimal invited set of expert ethical hackers and show how it scales with the adversarial threat. Fourth, we study the release-timing margin. Within the baseline model, feasibility of a BBP conditionally implies an earlier optimal release than in the no-BBP benchmark. We interpret this fourth result narrowly in practice: it is a within-model comparative static, and its practical relevance depends on operational readiness, triage throughput, and safe deployment capacity rather than following mechanically for every real-world BBP.

These results have practical implications. For managers, BBPs are best viewed as an additional governance layer rather than as a broad substitute for secure development practices. The baseline model highlights when monetary rewards versus reputational incentives are more effective, when severity-specific pricing is necessary to reduce exploitation risk, and when invite-set design matters in private programs. It also clarifies that the baseline model's conditional earlier-release margin is most plausible in practice when the commercial value of early launch remains steep, report-processing capacity is not swamped by low-quality or duplicative submissions, and valid reports can be translated into safely deployable fixes without substantial backlog. For policymakers and standard-setters, the results underscore the need to balance controlled-remediation benefits against transparency concerns, particularly when disclosure terms affect user awareness and precaution-taking. Together, these implications motivate a more integrated view of BBPs that links incentive design, disclosure governance, and release strategy.

## 2 Related literature

BBPs formalize post-release vulnerability discovery by rewarding ethical hackers (also known as white hat hackers or independent security researchers) for first, actionable reports under the BBP's rules of engagement. In this literature review, our focus is on production and operations management (POM) literature that connects security incentives to firms' operating and strategic choices (e.g., patching, disclosure, sourcing, cloud responsibility, release timing).[2] We complement this with targeted software security and BBP studies in computer science, information systems (IS), and economics that speak to bounty design, researcher behavior, disclosure, and platform dynamics. This blended view allows us to position a firm's *release timing* and *post-release SV discovery incentive design* (BBPs) inside an operations framework while engaging the specialized BBP evidence base that motivates our modeling assumptions and comparative statics.

### *2.1 Software security economics: Patching, disclosure, and reputational forces*

A core stream in POM and IS literature examines how liability, patching, and disclosure regimes shape firms' incentives for software security. Arora et al. (2006) model a software monopolist's trade-off between releasing earlier with more defects and investing in post-release patching. Because patching involves fixed costs and the marginal cost of distributing fixes is low, a larger market can make it optimal to *sell first, fix later*; they also show that the monopolist releases later and with fewer bugs than is socially optimal. Kim et al. (2011) analyze how product liability and patch-release policy interact to shape security investment and post-release patching behavior in a monopoly, helping explain when vendors rely more on post-release patching. Complementing this, Sen et al. (2020) study determinants of disclosure timing and document how producer characteristics and vulnerability features influence when discoverers choose to disclose, highlighting the strategic environment vendors face in the absence of formal coordination.

Empirical disclosure research connects public revelation to vendor behavior: Arora et al. (2010) show that disclosure *accelerates* patch release (i.e., increases the instantaneous patch-release rate, leading to shorter time-to-patch), consistent with disclosure imposing salient costs on vendors who delay. Theoretical work specifies when mediated disclosure is socially efficient. Cavusoglu et al. (2007) characterize responsible (coordinated) disclosure policies as those that minimize social loss and show non-obvious consequences of grace periods (the time window coordinators give vendors to develop patches before public disclosure) across single- and multi-vendor settings. Arora et al. (2008) model a social planner who sets disclosure timing to induce faster patches without unduly sacrificing quality, and analyze extensions with patch quality, partial adoption, and workarounds. Market-based mechanisms for vulnerability information also matter: Kannan and Telang (2005) compare unregulated market-based infomediaries against a public-good CERT-like coordinator and show that profit-driven infomediaries' leakage incentives can yield inferior welfare, motivating care in the design of market instruments such as BBPs.

On the organizational side, firms face both reputational pressures and operational challenges in managing security. Using a field quasi-experiment, Tang and Whinston (2020) show that reputational sanctions—publicly listing and ranking negligent entities—reduce security negligence, with measurable spillovers to unlisted firms. From a dynamic risk-management perspective, Bensoussan et al. (2020) model security that deteriorates both continuously and abruptly, offering structural guidance for adapting protection and monitoring under time-varying exposure. Likewise, Mookerjee and Samuel (2023) consider security control when vulnerabilities are only partially observable, aligning with our modeling of residual severe and non-severe vulnerabilities at release. Additional evidence shows sizable real consequences of breaches in healthcare settings (Kwon and Johnson, 2025) and highlights how network and audit policies shape social cost (Ghosh et al., 2025).

At a field level, Kumar and Mallipeddi (2022) survey cybersecurity in operations and supply chains and call for models that embed security choices into core operating and strategic decisions—precisely the integration we undertake by jointly endogenizing release timing and post-release incentives via a bounty mechanism. That need is underscored by Massimino et al. (2018), who document inattention to digital confidentiality in operations and supply chain research and advocate designs that treat security as first-order.

Responsible disclosure models provide microfoundations for when coordination and grace periods minimize social loss (Arora et al., 2008; Cavusoglu et al., 2007). Policy instruments outside the vendor also shape equilibrium risk: Png and Wang (2009) compare enforcement against attackers with facilitating user precautions and show the conditions under which each lowers expected losses more effectively. Together with evidence that disclosure hastens patching (Arora et al., 2010), these results motivate our explicit cost for uncoordinated public disclosure and the BBP's role in *internalizing* reports, structuring remediation, and shifting the vendor's calculus.

### *2.2 Software development, release timing, and product strategy*

We endogenize *release timing* jointly with a post-release discovery instrument. Related operations work offers complementary perspectives on pre/post-release effort. In maintenance, Kulkarni et al. (2009) formulate a queuing framework to optimally allocate effort, illuminating throughput–quality–delay trade-offs that also appear in pre-release testing. Coordination in distributed development affects the cost and timeliness of assurance; Xia et al. (2016) derive optimal coordination structures when development is distributed, implying that internal frictions can make pre-release testing more expensive or slower.

Crucially, Jiang et al. (2012) show that allowing *post-release testing* can rationalize earlier release yet lower lifetime field failures, even if more bugs remain at release, because testing continues in parallel with operations; their analysis also quantifies how market opportunity cost pulls release forward. On go-to-market levers, Li and Kumar (2022) analyze SaaS pricing and operations under subscription versus usage models, and Li et al. (2025) show how customizability and hybrid offerings change pricing and segmentation. Market structure also feeds back to quality choices: Zhou and Choudhary (2022) demonstrate how competition from open source can raise or lower proprietary quality and price depending on cost and usability differences.

**Additional related streams (e-Companion).** To keep the main paper focused on the operational mechanism and the analytical results, we relegate four complementary literature discussions to the e-Companion (Section EC.12). These cover (i) interdependence, sourcing, and cloud responsibility, clarifying how security risk propagates across organizational boundaries and how responsibility allocation shapes incentives; (ii) empirical and theoretical BBP research in adjacent fields, documenting researcher motivations, platform frictions, and program timing; (iii) contest-design analogues that motivate prize-setting and restricted-entry design in private BBPs; and (iv) policy-oriented analyses of interventions and disclosure governance that clarify when market mechanisms can substitute for, or require, regulatory support.

### 2.3 Positioning and contribution

While Arora et al. (2006) studied how patching could facilitate earlier releases, our setting is distinct in both question and mechanism. Rather than asking how patching technology alone rationalizes early release, we study how a *market design for post-release discovery*, a BBP, reshapes incentives. Specifically, we *jointly* endogenize (i) pre-release testing via release timing and (ii) post-release incentive design (bounties and access), in the presence of strategic ethical and malicious actors who compete in winner-take-all discovery. We characterize the parameter region in which a BBP exists (linking normalized illicit gains and reputational payoffs), derive *closed-form optimal bounties* for severe and non-severe findings, and show that for private programs the optimal invite set is strictly below, but increasing in, the expected number of adversaries. We also characterize a baseline release-timing result: within the baseline model, BBP feasibility conditionally implies earlier release relative to the no-BBP benchmark. Its practical applicability is narrower, because governed post-release discovery affects real firms only when they can support triage, validation, and safe deployment operationally. These channels operate through *who finds SVs first* and *how disclosures are controlled*, rather than through patching economics alone.

Relative to the extant literature on patching, disclosure (Ahmed et al., 2021), and interdependent risk, we provide a unified, micro-founded treatment that embeds BBPs within core operations choices. We contribute: (i) *closed-form bounty policies* that combine reputational payoffs, breach costs, and competition among researcher types, complementing patching/liability and disclosure-timing studies (Arora et al., 2008; Kim et al., 2011; Sen et al., 2020) and dynamic protection under partial observability (Bensoussan et al., 2020; Mookerjee and Samuel, 2023); (ii) a *feasibility region* for BBPs in terms of normalized illicit and reputational payoffs, clarifying when market mechanisms can substitute for (or need) policy support; (iii) a new design insight for private BBPs, namely that the *optimal invited expert set* is strictly below, but increasing in, the anticipated number of adversaries, aligning with restricted-entry contest rationales; and (iv) a characterization of the conditional release-timing margin in the baseline model. The last result is intentionally narrower than a claim that BBPs generally substitute for internal testing in practice. Formally, it is a within-model conditional implication; practically, it identifies when governed post-release discovery can alter the marginal calculus of delay while still leaving strong complementarity between BBPs and internal assurance in levels. Finally, our analysis complements adjacent BBP studies (see e-Companion Section EC.12.2) on researcher motivations, timing, and cost effectiveness by connecting those levers directly to a firm's *release decision* and quantifying the equilibrium reallocation of success probabilities between ethical and malicious hackers that underpins the profitability result and the baseline conditional timing result.

## 3 Model development

Software vendors face a recurrent release decision: ship sooner to capture time-sensitive commercial value and learning, or delay to reduce residual security risk. Pre-release software assurance includes manual and automated tests plus code review; it reduces, but does not eliminate, vulnerabilities. Virtually no complex software is released without residual vulnerabilities (Sen et al., 2020). Testing is costly in tools and skilled time, and it cannot replicate the full space of real user behavior or adversarial tactics. Adversarial techniques such as fuzzing help (Manès et al., 2019), yet no process can anticipate all cases. Post-release, firms combine several instruments to manage residual risk. Incident response and cybersecurity insurance primarily contain losses after exploitation. In contrast, patch management and BBPs aim to identify and remediate vulnerabilities before exploitation by accelerating discovery and controlled disclosure. Patching addresses known bugs once identified; post-release discovery occurs via internal monitoring, user reports, adversarial exploitation, public disclosure (e.g., CERT Coordination Center), and BBPs.

We next formalize vulnerability severity, since controlled disclosure choices and bounty design depend on impact. Software vulnerability severity reflects "the highest failure impact that the defect could cause" (IEEE Computer Society, 2010). One of the most widely used severity metrics, the Common Vulnerability Scoring System (CVSS), provides numerical scores (0.0–10.0) that correspond to qualitative scores: none, low, medium, high, and critical.[3] These ratings serve to inform users of potential impacts, help vendors prioritize fixes, and support vulnerability analysis (Munaiah and Meneely, 2016). We classify SVs into two categories: "severe" (corresponding to CVSS high and critical ratings) and "non-severe." These categories align with different hacker capabilities and incentives.[4]

### 3.1 Bug bounty programs

BBPs formalize post-release discovery: vendors publicly specify scope and rules and pay rewards to white-hat hackers for *first, valid, actionable* reports.[5] The bounty is paid only to the first valid reporter, so the mechanism is winner-take-all and pay-for-results. BBPs let vendors access specialized security expertise that would be costly to maintain in-house, while providing white-hat hackers with legal safe harbor, recognition, and monetary incentives. In contrast to fixed in-house testing budgets, BBP spending scales with delivered findings. By conditioning payment on private reporting and adherence to program rules, BBPs internalize reports and enable vendor-controlled disclosure until remediation. In practice, however, this governance function also requires verification, deduplication, severity classification, and researcher communication. These operating burdens matter for implementation because a program can attract low-quality or duplicative submissions, especially when report generation is partially automated, thereby raising effective triage cost even if gross discovery volume rises.

Large vendors sometimes run direct programs; smaller vendors frequently use platforms such as HackerOne or Bugcrowd (among others) that match researchers to programs, enforce safe-harbor norms, and provide triage and controlled-disclosure workflows. Platform-mediated BBPs have expanded participation and made programs viable for firms that could not attract sufficient independent attention. Our analysis focuses on *vendor-initiated* BBPs, which constitute the dominant model and align with our research question on vendor release and incentive design. Customer-initiated programs (e.g., an enterprise inviting testing of third-party software it relies on) involve different objectives and are outside our scope.

### 3.2 Model overview

Our model features a software vendor, white-hat hackers (WHHs), and black-hat hackers (BHHs). The vendor manages residual SV risk through release timing and severity-contingent bounties; hackers exert effort in winner-take-all discovery races. We distinguish expert WHHs (eWHH), who can discover severe and non-severe vulnerabilities, from non-expert WHHs (neWHH), who target non-severe vulnerabilities. BHHs operate outside legal bounds and pursue illicit gains by exploiting severe vulnerabilities (e.g., ransomware). We focus on technically sophisticated BHHs capable of high-impact exploitation; omitting less skilled adversaries who target only non-severe vulnerabilities centers the analysis on the policy-relevant risk of catastrophic loss. Types and motives are fixed during the game. Modeling gray-hat behavior or type switching would require a richer dynamic model and is left for future work.

We assume that $n$ eWHH, $l$ neWHH, and $m$ BHH simultaneously search for vulnerabilities. Hackers receive positive payoffs only if they are the first to report an SV to the software vendor. eWHHs receive both monetary rewards ($p_s$) and reputation gains[6] ($r_s$) when finding a severe SV first, giving a total reward ($r_s+p_s$).[7] For non-severe SVs, eWHH or neWHH receive only the monetary reward ($p_{ns}$) with no reputation gains. In practice, many BBPs announce bounties as ranges rather than fixed amounts, with the final payout determined by an ex-post evaluation of the vulnerability's impact. This introduces uncertainty for the hacker. However, we assume that hackers are rational actors who form expectations about the likely payout based on program rules and historical data. In our model, the fixed points $p_s$ and $p_{ns}$ represent the expected monetary rewards for the respective vulnerability categories. Assuming risk-neutral hackers (as implied by the expected payoff maximization in equation (8)), the expected value is the appropriate driver of effort optimization. This simplification enhances tractability without losing the essence of the strategic dynamics.

When BHHs find a severe SV before the rest, they obtain an illicit payoff ($W$), which we interpret as the *ex ante expected* monetization conditional on first discovery. This expectation aggregates feasible monetization channels (e.g., ransomware or extortion payments, data theft and sale, and sale of a zero-day exploit) and the likelihood of successful monetization. Ransomware monetization and ransomware-as-a-service arrangements can raise this expectation by enabling scalable deployment and revenue sharing, but we treat $W$ as a reduced-form parameter rather than making empirical claims about its level. In e-Companion Section EC.8, we allow $W$ to be stochastic; under risk neutrality, incentives depend on $\mathbb{E}[W]$, so the main pricing and feasibility implications carry through with $W$ interpreted as this expectation. In our baseline analysis, we take participation levels $(n, l, m)$ as exogenous and treat the reputational payoff $r_s$ as a primitive in the baseline in order to focus on effort choices and the vendor's monetary policy. Section 3.11 optimizes $n$ for private programs. In the e-Companion, Section EC.11 extends the model by endogenizing participation through outside options and heterogeneous expertise.

BBP design affects participation and the skill mix of participating researchers. To incorporate this channel, e-Companion Section EC.11 introduces heterogeneous expertise among expert WHHs and an increasing outside option, yielding a cutoff participation condition and an endogenous expected expertise level among participants. In that extension, the severe bounty $p_s$ affects outcomes through both an effort channel and a composition channel (higher $p_s$ attracts higher-expertise participants), bringing the model closer to the participation dynamics emphasized in practice.

E-Companion Table EC.1 provides abbreviations and variable definitions.

## 3.3 Hacker efforts, costs, and payoffs

Hackers choose their effort levels to find SVs, incurring quadratic costs. Let $\alpha_{is} \in (0, 1), \mu_{is} \in (0, 1)$ be the effort of the $i$-th eWHH and BHH, respectively, to find severe SVs and let $\alpha_{ins}, \beta_{ins}$ be the effort of the $i$-th eWHH and neWHH to find non-severe SVs. The cost function for eWHHs accounts for their ability to work on both severe and non-severe SVs. This cost is given by the following equation:

$$F_{ie} = \frac{c_w \alpha_{is}^2}{2} + \frac{\alpha_{ins}^2}{2} + \alpha_{is}\alpha_{ins}, \quad \text{where } c_w > 1 \qquad (1)$$

Throughout, the effort variables should be interpreted as *search intensity* (the intensity or share of attention devoted to searching for a class of vulnerabilities) rather than as a sequential count of distinct bugs. The convexity of the cost function is a reduced-form way to capture increasing marginal opportunity and coordination costs of raising intensity (e.g., attention, verification, and reporting bandwidth), which helps support an interior equilibrium in the winner-take-all discovery contest. This interpretation does not require that the "next" vulnerability is technologically harder than the previous one.

The parameter $c_w$ serves as a severity-adjusted effort cost multiplier, reflecting the additional complexity and resource demands of discovering severe SVs compared to non-severe SVs. The interaction term, $\alpha_{is} \times \alpha_{ins}$, captures the trade-off in effort due to the hacker's finite resources (such as time and cognitive bandwidth). This standard economic formulation for multi-tasking implies that as an eWHH increases effort on severe SVs, the marginal cost of working on non-severe SVs rises, and vice versa. This reflects increased opportunity costs and the relative efficiency of specialization, rather than a technical dependency between the vulnerabilities themselves. As neWHHs lack the expertise to discover severe SVs, they allocate their effort solely to non-severe SVs, thus avoiding the extra cost of working on complex vulnerabilities and simplifying the cost function of neWHHs:

$$F_{ine} = \frac{\beta_{ins}^2}{2} \qquad (2)$$

BHHs exclusively target severe SVs, as non-severe SVs offer little illicit gains. It is important to distinguish the BHH's expected illicit gain ($W$) from the vendor's total cost of a severe breach ($TC_s$, introduced later). While $W$ and $TC_s$ may be correlated, they are conceptually distinct: $W$ reflects attacker-side monetization, whereas $TC_s$ encompasses the vendor's total losses, including remediation expenses, regulatory penalties, and reputational damage, which may substantially exceed $W$. Our model treats them as independent parameters to analyze how attacker incentives ($W$) versus vendor losses ($TC_s$) independently influence the vendor's strategy. BHHs' cost function is expressed as follows:

$$F_{ib} = \frac{c_b \mu_{is}^2}{2}, \text{ where } c_b > 1 \qquad (3)$$

The parameters $c_w$ and $c_b$ can also be interpreted as reduced-form shifters of the cost of generating search intensity under the prevailing tool environment. Improvements in automation and AI can lower these effective costs, changing equilibrium effort and success probabilities through the same incentive channels as monetary and reputational rewards. Importantly, our effort variables represent *search intensity* in a time-bound race, not a sequential count of distinct bugs. Thus, even if AI lowers baseline scanning or coding cost, scaling up intensity still requires verification, deduplication, exploit development, and coordination, so increasing marginal costs need not disappear. This point is especially relevant when AI-assisted tools generate large volumes of baseline anomalies or low-quality reports: the bottleneck shifts toward human cognitive bandwidth and triage rather than vanishing altogether. If AI lowers $c_w$ and $c_b$ symmetrically, the discovery race accelerates but the basic competitive logic remains unchanged. If it lowers $c_b$ more than $c_w$, then the normalized illicit-gain term $W/c_b$ rises relative to the normalized reputational term $r_s/c_w$, which increases the vendor's optimal severe bounty and tightens the feasibility conditions for shifting first discovery from malicious to ethical hackers. E-Companion Section EC.7 formalizes robustness to alternative convex effort costs and AI-driven changes in effective cost levels.

## 3.4 Release time and the likelihood of residual SVs

Software vendors face a critical trade-off: launching products rapidly to capture time-sensitive commercial value, customer learning, and ecosystem positioning versus delaying release to conduct rigorous pre-release security testing. The release time, $t$, is the moment a vendor launches the software after testing; launching too early risks undiscovered vulnerabilities, while extended testing reduces vulnerabilities but can lead to lost commercial value. Let $R(t)$ denote the vendor's expected lifetime launch value as a function of $t$, where $R'(t) < 0$, so that delay of release reduces value. This reduced-form term can reflect foregone sales, delayed customer learning, postponed integration benefits, or other time-to-market losses. If changes in the software environment, including stronger AI-mediated automation or interface bypass, flatten this launch-value profile, then the timing incentives studied below become weaker even if the BBP's discovery-allocation effect remains. Furthermore, let $K_s(t)$ and $K_{ns}(t)$ represent the probability that a severe or non-severe vulnerability remains at release time $t$ after the vendor's internal security pipeline. They summarize layered in-house software security controls such as secure-by-design practices, automated testing and scanning in Continuous Integration and Continuous Deployment, code review, and penetration testing; releasing later captures additional internal assurance that reduces residual vulnerabilities. The BBP is modeled as an additional post-release layer that governs external discovery and disclosure of any residual vulnerabilities that remain at launch.

We assume $K'_s(t) < 0$ and $K'_{ns}(t) < 0$, implying that longer testing helps vendors fix more vulnerabilities, and $K''_s(t) > 0$ and $K''_{ns}(t) > 0$, reflecting the diminishing returns to testing. Finally, let $\delta \in [0, 1]$ denote the probability that, conditional on a residual non-severe vulnerability existing at release and not being found first by a WHH, users discover it during the relevant timeframe. We model $K_s(t)$ and $K_{ns}(t)$ as smooth functions for tractability. In practice, internal testing may be lumpy (e.g., distinct phases of automated scanning, code review, and penetration testing). In such cases, $K(t)$ should be interpreted as the lower envelope of residual vulnerability levels achievable by an optimally scheduled sequence of testing activities. Our key results rely on the property that extending pre-release assurance reduces expected residual risk with diminishing returns on average, a property that holds for the envelope of phased testing.

## 3.5 Success probabilities in vulnerability discovery

For any type of hacker, success is defined as being the first to discover a software vulnerability (SV). Success in vulnerability discovery is inherently a winner-take-all competition.[8] For WHHs, this structure emerges naturally from BBP rules that reward only the first WHH to report a previously unknown vulnerability. In general, being second yields no reward, regardless of independent discovery or effort invested. Similarly, while multiple BHHs could theoretically benefit from exploiting the same vulnerability, practical dynamics often make it a winner-take-all scenario. Once a BHH exploits a vulnerability and successfully extracts illicit gains, the exploitation itself often alerts the software vendor, leading to emergency workarounds or patching that prevents further exploitation by other BHHs.[9]

To make the analysis tractable, we focus on type-symmetric equilibria where hackers of the same type (eWHH, neWHH, and BHH) face similar payoff structures and choose the same effort levels. In a type-symmetric equilibrium, the probability that any hacker discovers a bug first is inversely proportional to the total number of competitors—$(n + m)$ for severe vulnerabilities and $(n + l)$ for non-severe vulnerabilities. Allowing for a single hacker to deviate from the type symmetric outcome, we nuance the base success probability to depend on the deviating hacker's relative effort compared to competitors. Consider a focal eWHH "$i$" who invests effort $\alpha_{is}$ in finding severe SVs. The average effort of the other $(n - 1)$ eWHHs and $m$ BHHs who devote effort to finding severe SVs is given by $(n-1)\alpha_s + m\mu_s/n+m-1$, where $\alpha_s$ and $\mu_s$ are the effort levels selected by the two types at the type-symmetric equilibrium. We assume that when the focal eWHH $i$, deviates from the type symmetric equilibrium to increase her effort, her probability of finding the bug increases by the difference between her effort level $\alpha_{is}$ and the average effort allocated by all other hackers searching for severe bugs, $(n-1)\alpha_s + m\mu_s/n+m-1$. Thus, the probability for eWHH $i$ to find a severe bug first is as follows:

$$\mathbb{P}^s_{ie} = \frac{1}{n+m} + \frac{1}{n+m}\left(\alpha_{is} - \frac{(n-1)\alpha_s + m\mu_s}{n+m-1}\right) \tag{4}$$

Similarly, the success probabilities for eWHH $i$ to find a non-severe SV ($\mathbb{P}^{ns}_{ie}$), for neWHH $i$ to find a non-severe SV ($\mathbb{P}^{ns}_{ine}$), and for BHH $i$ to find a severe SV ($\mathbb{P}^{s}_{ib}$), are given, respectively, by the following equations:

$$\mathbb{P}^{ns}_{ie} = \frac{1}{n+l} + \frac{1}{n+l}\left(\alpha_{ins} - \frac{(n-1)\alpha_{ns} + l\beta_{ns}}{n+l-1}\right) \tag{5}$$

$$\mathbb{P}^{ns}_{ine} = \frac{1}{n+l} + \frac{1}{n+l}\left(\beta_{ins} - \frac{n\alpha_{ns} + (l-1)\beta_{ns}}{n+l-1}\right) \tag{6}$$

$$\mathbb{P}^{s}_{ib} = \frac{1}{n+m} + \frac{1}{n+m}\left(\mu_{is} - \frac{n\alpha_s + (m-1)\mu_s}{n+m-1}\right) \tag{7}$$

*3.5.1 Exponential and Weibull distributed discovery time.* Note that our assumed functional form for the success probabilities is qualitatively similar to that derived from an exponential distribution of bug discovery times (see e-Companion Section EC.2). In the exponential model, the probability that the focal eWHH $i$ finds a bug before time $\tau$ is given by $F(\tau; \alpha_i) = 1 - \exp(-\alpha_i \tau)$ and that each BHH $j$ finds a bug before time $\tau$ is given by $F(\tau; \mu_j) = 1 - \exp(-\mu_j \tau)$. With such a formulation, the probability that eWHH $i$ finds a bug first can be derived as $\alpha_{is}/(\alpha_{is} + (n-1)\alpha_s + m\mu_s)$. Hence, the expression for the success probability, in this case, is decreasing with $n$, $m$, $\alpha_s$, and $\mu_s$ and increasing in $\alpha_{is}$. The same comparative statics are preserved in our simplified linear and additive specification. While the exponential model provides a more theoretically grounded approach to vulnerability discovery times, our simpler formulation preserves the essential competitive dynamics among the different types of hackers while providing greater analytical tractability. Furthermore, we show that our main comparative statics are robust to an increasing-hazard Weibull specification (e-Companion Section EC.4). These assumptions yield closed-form expressions for (i) the probability that ethical versus malicious hackers are first, (ii) equilibrium effort responses, and (iii) optimal severity-contingent bounty design and release timing. We emphasize that these assumptions are not intended as literal empirical claims about discovery-time distributions; rather, they provide a disciplined baseline for studying how BBPs shift the race for *first discovery* and how this interacts with residual vulnerabilities $K(t)$ and time-to-market value $R(t)$.

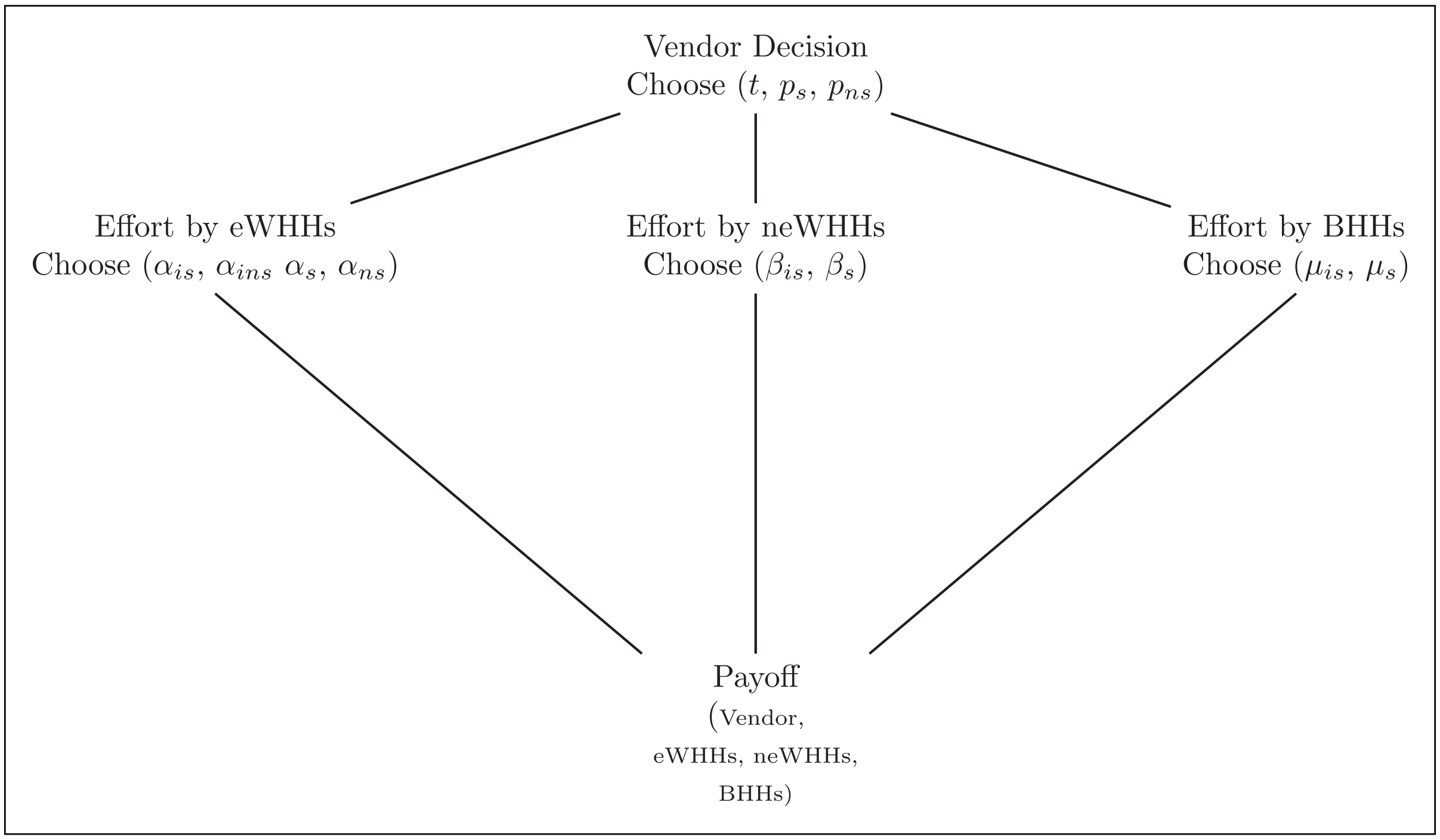

**Figure 1.** Sequence of decisions in the Stackelberg game: The vendor chooses release time $t$ and bounties $(p_s, p_{ns})$; after release, ethical and malicious hackers choose search efforts.

### 3.6 *Two-stage Stackelberg game formulation*

We model the strategic interaction between the software vendor and hackers as a Stackelberg game in which the vendor acts as the leader. Figure 1 depicts the sequence of the game. In the first stage, the software vendor commits to software release timing $t$ and bounty amounts $p_s$ and $p_{ns}$ for severe and non-severe SVs, respectively. These commitments become common knowledge for all. In the second stage, after observing the vendor's choices, multiple hackers engage in a simultaneous-move subgame where they choose their effort levels in finding vulnerabilities. Specifically, expert WHHs choose effort $\alpha_{is}$ and $\alpha_{ins}$, non-expert WHHs choose effort $\beta_{ins}$, and BHHs choose effort $\mu_{is}$.

We solve this two-stage game for subgame perfect equilibrium using backward induction. The solution involves first deriving the Nash equilibrium of the second-stage subgame among hackers for a fixed bounty amount and release time by the vendor in stage one. Then, incorporating these anticipated hacker responses, we solve the vendor's first-stage optimization problem to determine the optimal release timing and bounty amounts.

### 3.7 *Optimal effort of the hackers*

In the second stage, hackers respond to the vendor's choices of release time $t$ and bounty amounts $p_s$ and $p_{ns}$ by choosing their optimal effort levels. The total expected payoff of the focal eWHH $i$, $R_{ie}$, is given by the following equation:

$$R_{ie} = K_s(t)\mathbb{P}^s_{ie}(r_s + p_s) + K_{ns}(t)\mathbb{P}^{ns}_{ie}(p_{ns}) - F_{ie} \tag{8}$$

The first and second terms on the RHS of (8) are the expected payoffs from finding severe and non-severe SVs, respectively. The last term is the effort cost of finding the SVs. Substituting equations (4) and (5) for success probabilities and equation (1) for the effort costs in (8) leads to the following equation:

$$\begin{aligned} R_{ie} = {} & K_s(t)\frac{1}{n+m}\left[1 + \left(\alpha_{is} - \frac{(n-1)\alpha_s + m\mu_s}{n+m-1}\right)\right](r_s + p_s) \\ & + K_{ns}(t)\frac{1}{n+l}\left[1 + \left(\alpha_{ins} - \frac{(n-1)\alpha_{ns} + l\beta_{ns}}{n+l-1}\right)\right](p_{ns}) \\ & - \left[\frac{c_w\alpha_{is}^2}{2} + \frac{\alpha_{ins}^2}{2} + \alpha_{is}\alpha_{ins}\right] \end{aligned} \tag{9}$$

The total expected payoff for neWHH is given by the following equation:

$$\begin{aligned} R_{ine} &= K_{ns}(t)\,\mathbb{P}^{ns}_{ine}(p_{ns}) \;-\; F_{ine} \\ R_{ine} &= K_{ns}(t)\frac{1}{n+l} \\ & \quad \times\left[1 + \left(\beta_{ins} - \frac{n\alpha_{ns} + (l-1)\beta_{ns}}{n+l-1}\right)\right](p_{ns}) - \frac{\beta_{ins}^2}{2} \end{aligned} \tag{10}$$

Finally, the total expected payoff for BHH is given by, where $W$ denotes the *ex ante expected* illicit payoff conditional on being first to discover a severe vulnerability:

$$R_{ib} = K_s(t)\,\mathbb{P}^s_{ib}(W) \; - \; F_{ib}$$

$$R_{ib} = \frac{K_s(t)W}{n+m}\left[1 + \left(\mu_{is} - \frac{n\alpha_s + (m-1)\mu_s}{n+m-1}\right)\right] - \frac{c_b\mu_{is}^2}{2} \tag{11}$$

In a type-symmetric equilibrium, the eWHH may allocate effort to discovering severe and non-severe SVs (i.e., $\alpha_{is} > 0$ and $\alpha_{ins} > 0$) or to discovering only severe bugs (i.e., $\alpha_{is} > 0$ and $\alpha_{ins} = 0$). While eWHHs may work on discovering both severe and non-severe vulnerabilities, anecdotally, the second case seems more prevalent among expert bounty hunters circa 2024. Expert bounty hunters are not only attracted by the higher rewards for severe SVs but also by the challenge of finding technically complex hacking attacks. Additionally, severe SVs are more likely to get adjudicated swiftly by BBP for bounty rewards. Thus, we relegate the analysis of the first case (i.e., $\alpha_{is} > 0$ and $\alpha_{ins} > 0$) to e-Companion Section EC.3 and focus our discussion in the paper on the more important, second case (i.e., $\alpha_{is} > 0$ and $\alpha_{ins} = 0$). Hacker behavior in the type-symmetric Nash equilibrium is reported in Lemma 1.

**Lemma 1** (Hackers' Optimal Effort)**.** *When $K_s(t)(r_s + p_s)/(n+m)c_w > K_{ns}(t)p_{ns}/n + l$, $\alpha_{is} > 0$, and $\alpha_{ins} = 0$. For given bounty amounts ($p_s$, $p_{ns}$) and release time $t$ selected by the software vendor, hackers choose the following optimal effort levels at the type-symmetric equilibrium:*

$$\alpha_{is} = \alpha_s = \frac{1}{(n+m)c_w}K_s(t)(r_s + p_s), \quad \alpha_{ins} = 0 \tag{12}$$

$$\beta_{ins} = \beta_{ns} = \frac{1}{l}K_{ns}(t)(p_{ns}) \tag{13}$$

$$\mu_{is} = \mu_s = \frac{1}{c_b(n+m)}K_s(t)(W) \tag{14}$$

Lemma 1 states that expert white hats do not allocate any effort to finding non-severe SV when the expected payoff of eWHH from severe bugs (normalized by the effort cost multiplier) is greater than the expected payoff from non-severe bugs. Additionally, it specifies the optimal effort allocation for each type of hacker. These optimal solutions follow directly from the concavity of hackers' payoff functions with respect to their effort choices ($\alpha_{is}, \alpha_{ins}, \beta_{ins}$, and $\mu_{is}$), which ensures that the first-order conditions yield global maxima. At the symmetric equilibrium, these conditions produce the optimal effort levels shown in equations (12) to (14). Examining these expressions, we note that higher likelihood of residual SVs motivates hackers to increase their search efforts, as the probability of successful discovery increases. The expected rewards also play a crucial role, whether through bounties and reputation gains for ethical hackers or illicit gains for malicious actors, with larger potential payoffs driving increased effort allocation. The competitive dynamics among hackers manifest themselves through an inverse relationship between the effort levels and the number of competing hackers looking for the same vulnerability type. When more hackers are searching for a particular vulnerability category, individual hackers reduce their effort investment, reflecting the decreased probability of being the first to discover a vulnerability in a crowded field. Finally, severity-adjusted effort cost multipliers act as moderating factors, with lower costs encouraging greater effort allocation across all types of hackers.

The relationship between these effort levels and discovery outcomes is captured in Lemma 2, which derives equilibrium first-discovery probabilities for severe vulnerabilities (expert WHHs vs. attackers) and a reduced-form non-severe reporting yield for non-expert WHHs in the type-symmetric equilibrium.[10]

**Lemma 2** (Hackers' Success Probabilities)**.** *For fixed bounty amounts ($p_s$ and $p_{ns}$) and release time $t$, the following success probabilities emerge at the type-symmetric equilibrium based on hackers' optimal effort choices:*

$$\mathbb{P}^s_{ie} = \max\left(0,\; \frac{1}{n+m}\left[1 + \frac{mK_s(t)}{(n+m-1)(n+m)} \times \left\{\frac{(r_s+p_s)}{c_w} - \frac{W}{c_b}\right\}\right]\right) \tag{15}$$

$$\mathbb{P}^{ns}_{ine} = \frac{K_{ns}(t)p_{ns}}{l} \tag{16}$$

$$\mathbb{P}^s_{ib} = \max\left(0,\; \frac{1}{n+m}\left[1 + \frac{nK_s(t)}{(n+m-1)(n+m)} \times \left\{\frac{W}{c_b} - \frac{(r_s+p_s)}{c_w}\right\}\right]\right) \tag{17}$$

The severe-vulnerability success probabilities in (15) and (17) are obtained by substituting the optimal effort levels (equations (12) to (14)) into the severe-vulnerability probability expressions (equations (4) and (7)) and imposing type symmetry.[11] For each type of hacker, the probabilities reveal key insights into competitive dynamics. For eWHH, the success probability increases with a higher reward-to-cost ratio and the likelihood of residual vulnerabilities, while decreasing with BHHs' illicit gain-to-cost ratio. The neWHHs' success probability exhibits a simpler relationship, being directly proportional to both the likelihood of residual non-severe vulnerabilities and the bounty offered, while inversely related to the number of competing neWHHs. BHHs' success probability, conversely, increases with their illicit gain-to-cost ratio and decreases with the ethical hackers' reward-to-cost ratio, while also being positively related to the likelihood of residual vulnerabilities. These probabilities underscore the competitive nature of vulnerability discovery between ethical and malicious hackers. The success probability of one group necessarily affects the other in a winner-take-all dynamic, where only the first to discover a vulnerability receives the benefit. This relationship is particularly evident in severe vulnerabilities, where eWHHs and BHHs directly compete for discovery. The relationship between bounty amounts and success probabilities is particularly notable, as illustrated in Figure 2. As the monetary reward (bounty) offered to WHHs increases, their success probability rises, while that of BHHs decreases. The increased success probability for WHHs stems from higher bounties incentivizing them to exert more effort in searching for vulnerabilities, improving their chances of discovery before BHHs can exploit them.

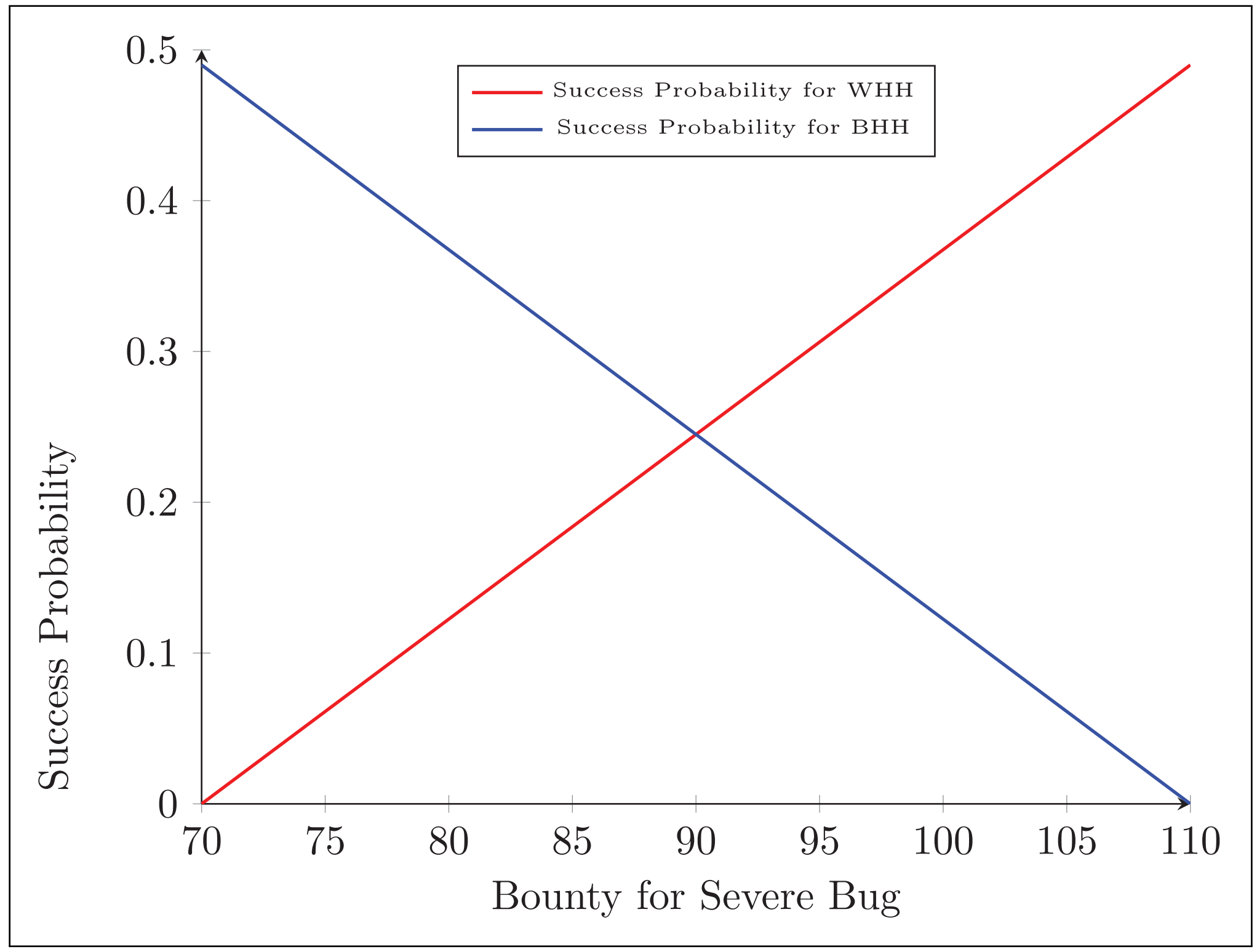

**Figure 2.** Success probability for WHH and BHH versus bounty for SV offered to WHH. WHH = white-hat hacker; BHH = black-hat hacker; SV = software vulnerability.

The timing of software release plays a crucial role through its effect on the likelihood of residual vulnerabilities. Earlier releases typically mean higher likelihood of residual vulnerabilities, which increases the success probabilities for all types of hackers proportionally. This relationship between release timing and vulnerability discovery success probabilities becomes particularly important when considering the vendor's optimal release strategy.

### 3.8 Software vendor's optimal bounty strategy

In the first stage of the game, the software vendor determines the bounty amounts $p_s$ and $p_{ns}$ along with the release time $t$. We designate by $TC_s$ the vendor's cost if a severe SV occurs and a BHH succeeds in finding it first. We further designate by $TC_{ns}$ the vendor's cost if a non-severe SV occurs and no WHH finds it, but rather a user finds it, where $TC_s \gg TC_{ns}$. We interpret $TC_s$ as the present value of the *total* loss to the vendor if a severe vulnerability is found first by a BHH and exploited. This loss includes incident response and remediation (including patch development and deployment), regulatory and legal costs, reputational harm, and loss of future demand. In particular, for subscription (SaaS) business models, $TC_s$ may include expected customer churn and the associated loss of future recurring revenue; we make this channel explicit, along with a remediation-capability extension, in the e-Companion (Section EC.9). The profit function of a software vendor participating in a BBP is given by the following equation[12] :

$$
\begin{aligned}
\Pi = R(t) - K_s(t)\left(m\mathbb{P}^s_{ib}\right) TC_s \quad &\text{(2nd term)} \\
- K_s(t)\left(n\mathbb{P}^s_{ie}\right) p_s \quad &\text{(3rd term)} \\
- K_{ns}(t)\left(l\mathbb{P}^{ns}_{ine}\right) p_{ns} \quad &\text{(4th term)} \\
- \delta\, K_{ns}(t)\, TC_{ns} \left[1 - l\mathbb{P}^{ns}_{ine}\right] \quad &\text{(5th term)}
\end{aligned} \qquad (18)
$$

Note that for the cost terms (i.e., the second, third, fourth, and fifth RHS terms), the individual success probabilities of hackers are scaled by the number of hackers of that type, reflecting the fact that the vendor incurs a cost if any one of these hackers discovers the SV. Term 2 is the cost incurred by the vendor when one of the BHHs finds the vulnerability first, and term 3 is the cost incurred in paying bounties for severe bugs. Cost term 4 is the cost incurred for paying bounties for non-severe bugs, and cost term 5 relates to non-severe SVs discovered by software users (not hackers) with probability $\delta$, requiring the vendor to fix the SVs. Plugging in the expressions for success probabilities into the vendor's objective function, we obtain the following equation:

$$\begin{aligned} A(t) &\equiv \frac{K_s(t)}{(n+m-1)(n+m)} \\ \Pi &= R(t) - K_s(t)\frac{m}{n+m} \\ &\quad \times \left[1 + nA(t)\left\{\frac{W}{c_b} - \frac{(r_s+p_s)}{c_w}\right\}\right](TC_s) \\ &\quad - K_s(t)\frac{n}{n+m}\left[1 + mA(t)\left\{\frac{(r_s+p_s)}{c_w} - \frac{W}{c_b}\right\}\right](p_s) \\ &\quad - (K_{ns}(t)p_{ns})^2 - \delta K_{ns}(t)TC_{ns}\left[1 - K_{ns}(t)p_{ns}\right] \end{aligned} \tag{19}$$

Recall that $R(t)$ decreases with delay ($R'(t) < 0$). Moreover, we assume that further postponement leads to a progressively steeper decline in launch value (i.e., $R''(t) < 0$). It is crucial to note that $R(t)$ represents the total expected lifetime launch value contingent on the launch date $t$, not the trajectory of revenue over calendar time post-launch. The assumption $R''(t) < 0$ can reflect accelerating losses from delayed customer adoption, delayed learning, postponed complementor integration, contractual launch commitments, or worsening competitive position. Mathematically, this assumption, combined with $K''(t) > 0$, serves as a sufficient condition to ensure the concavity of the objective function. Such convex delay penalties can also reflect operational frictions such as integration and regression testing, deployment coordination with customers and partners, and market-facing launch commitments. If technological change or AI-mediated disintermediation flattens launch-value gradients, the conditional earlier-release effect derived below weakens or can disappear even though the bounty-design and first-finder reallocation results remain.

To characterize the vendor's optimal strategy, we first analyze the optimal bounty choices for any chosen release timing decision. The following proposition characterizes the optimal bounties $p_s$ and $p_{ns}$ offered by the software vendor:

**Proposition 1** (Optimal Bounty Amounts)**.** *Conditional on the vendor's release-time choice, the optimal bounty for severe vulnerabilities balances (i) the vendor's expected marginal loss from a severe breach and the malicious hackers' expected illicit gains against and (ii) the vendor's expected payout cost and the incentives needed to shift discovery toward ethical researchers. Consequently, the optimal severe-vulnerability bounty is increasing in the vendor's loss from a severe breach and in malicious hackers' illicit gains, and decreasing in ethical hackers' reputational incentives. The optimal bounty for non-severe vulnerabilities is proportional to the expected loss from non-severe defects that are discovered by users.*

**Characterization (closed-form expressions).** The optimal bounty levels are given by the following equations:

$$p_s = \frac{1}{2}\left[TC_s + \frac{c_w}{c_b}W - r_s\right] - \frac{1}{2}\frac{(m+n)(m+n-1)}{m}\frac{1}{K_s(t)}c_w \tag{20}$$

$$p_{ns} = \frac{\delta\, TC_{ns}}{2} \tag{21}$$

*Interpretation.* Equation (20) shows how the severe-bug bounty internalizes breach losses and malicious incentives while netting out reputational incentives and the cost of shifting the discovery race. Equation (20) also shows that the optimal severe bounty is increasing in attacker monetization, with

$$\frac{\partial p_s^*}{\partial W} = \frac{c_w}{2c_b} > 0$$

Thus, higher expected attacker monetization requires higher posted bounties to shift first discovery toward ethical reporting. In practice, this mapping suggests using attacker-monetization indicators as inputs to internal estimates of $W$ when setting severe-vulnerability bounty levels.

When eWHHs differ in expertise and participation is endogenous, higher $(p_s + r_s)$ can also improve outcomes by attracting a higher-expertise participating pool; e-Companion Section EC.11 formalizes this composition (screening) channel. The vendor's profit function (equation (19)) is concave in $p_s$ and $p_{ns}$, implying that the first-order conditions are both necessary and sufficient for determining the optimal bounty amounts given in Proposition 1 (see e-Companion Section EC.5). These optimal bounties illustrate how vendors balance security investments between pre- and post-release phases. For severe vulnerabilities, the optimal bounty $p_s$ increases with potential exploitation costs ($TC_s$) and BHHs' normalized illicit gains ($W/c_b$), decreases with the reputation gains of white hat hackers ($r_s$)—as these intrinsic rewards partially substitute for monetary compensation—and rises with the likelihood of residual vulnerabilities at release ($K_s(t)$). In contrast, for non-severe vulnerabilities, the optimal bounty equals one half of the *expected* user-discovery cost, $p_{ns} = \delta\, TC_{ns}/2$, reflecting simpler risk considerations when only WHHs are involved.

The closed-form bounty schedule provides a micro-founded pricing rule for coordinated post-release discovery that complements operations work on patching/liability and disclosure timing. By jointly incorporating vendor breach costs, attackers' normalized illicit gains, and ethical hackers' reputational payoffs, the policy links incentive design to the speed–security trade-off emphasized in the OM and IS literature (see e-Companion Section EC.12.4). It also connects BBPs to the economics of innovation tournaments, where prize magnitude and access rules shape participation and effort (see e-Companion Section EC.12.3). The severe-bounty formula clarifies how monetary rewards should rise with breach stakes and attackers' outside options and fall with the strength of reputational payoffs. Because the optimal severe bounty increases with the likelihood of residual severe vulnerabilities at release, the design is intrinsically dynamic: as testing reduces residual risk, the efficient severe bounty declines, tying payout policy to the vendor's release timing and test intensity choices. Managerially, the rule has two practical implications. First, bounty budgets should be calibrated against two observables: internal breach-cost models for severe incidents and external indicators of exploit value or attacker effort cost, which jointly proxy the attacker's normalized gains. Second, vendors can economize on monetary outlays by deliberately cultivating reputation mechanisms, public acknowledgments, leaderboard visibility, and common vulnerabilities and exposures (CVE) credits, because reputational benefits substitute for cash in the optimal policy. Together, these design principles translate the theory into a compensation scheme that is consistent with controlled disclosure objectives in operations and with observed contest dynamics, and they situate BBP pricing squarely within core OM decisions on quality assurance and post-release remediation (see e-Companion Sections EC.12.3 and EC.12.4).

*3.8.1 Reputational incentives as a vendor design lever.* In the model, $r_s$ captures non-cash rewards to expert WHHs (e.g., public recognition, leaderboard visibility, CVE credit, and career signaling). Vendors can influence these reputational benefits through program design. Proposition 1 implies a direct cash-versus-reputation substitution:

$$\frac{\partial p_s^*}{\partial r_s} = -\frac{1}{2} < 0$$

so increasing recognition reduces the required cash bounty by one-half at the margin (i.e., $dp_s^* = -\frac{1}{2}\, dr_s$). To incorporate that recognition is not free, we interpret enhancing $r_s$ as requiring an implementation cost (e.g., triage quality, public acknowledgments, compliance and legal safe-harbor efforts). This highlights an operational trade-off: vendors can reduce cash payouts by investing in credible reputation mechanisms that increase researcher expected utility.

## 3.9 Feasibility region for the existence of a BBP

For a bounty program to exist, the vendor should be willing to pay positive bounties to WHHs. In addition, it should also be the case that WHHs and BHHs have positive probabilities of finding SVs first. Using the expressions we derive for the success probabilities in equations (15) to (17) and the optimal bounties in equations (20) to (21), we can obtain conditions to support the existence of a bounty program. Lemma 3 reports the requirements the parameters should satisfy.

**Lemma 3** (Feasibility of a BBP)**.** *A BBP is feasible, in the sense that the vendor optimally offers a strictly positive severe-vulnerability bounty and the severe-bug discovery contest yields interior success probabilities for both ethical and malicious hackers, if and only if the net incentive advantage of malicious hacking (normalized illicit gains) over ethical hacking (normalized reputational incentives) lies within a bounded interval. The lower and upper bounds of this interval are determined by the vendor's breach losses, the intensity of competition in the severe-bug discovery contest (the number of ethical and malicious researchers searching for severe bugs), and the release-time-dependent severity exposure.*

**Characterization (feasibility interval).** A BBP exists (i.e., $p_s > 0$, $\mathbb{P}^s_{ie} > 0$, and $\mathbb{P}^s_{ib} > 0$) if and only if

$$LB < \left[\frac{W}{c_b} - \frac{r_s}{c_w}\right] < UB \tag{22}$$

where the bounds are defined as follows:

$$LB \equiv \max\{LB_1, LB_2\} \tag{23}$$

$$LB_1 \equiv \frac{(m+n)(m+n-1)}{m\,K_s(t)} - \frac{TC_s}{c_w}$$

$$LB_2 \equiv \frac{TC_s}{c_w} - \frac{(2m+n)(m+n)(m+n-1)}{mn\,K_s(t)} \tag{24}$$

and

$$UB \equiv \frac{(m+n)(n+m-1)}{m\,K_s(t)} + \frac{TC_s}{c_w} \tag{25}$$

Figure 3 also helps interpret how attacker monetization enters the feasibility interval in Lemma 3. The key object is the net normalized incentive gap $\left[\frac{W}{c_b} - \frac{r_s}{c_w}\right]$ in equation (22). Here $W$ should be interpreted as the ex ante expected monetization of a particular severe vulnerability conditional on first discovery. In the baseline model, $W$ is therefore treated as an exogenous expectation anchored by practical exploitability and target value, rather than as a direct function of market-wide zero-day scarcity. If one nevertheless expects scarcity to raise this expected monetization at the margin, then equation (20) implies that the vendor must post a higher severe bounty to remain competitive, and equation (22) implies that feasibility tightens as the normalized attacker advantage rises. E-Companion Section EC.8 captures related uncertainty by allowing $W$ to be stochastic while preserving the main pricing and feasibility logic under risk neutrality.

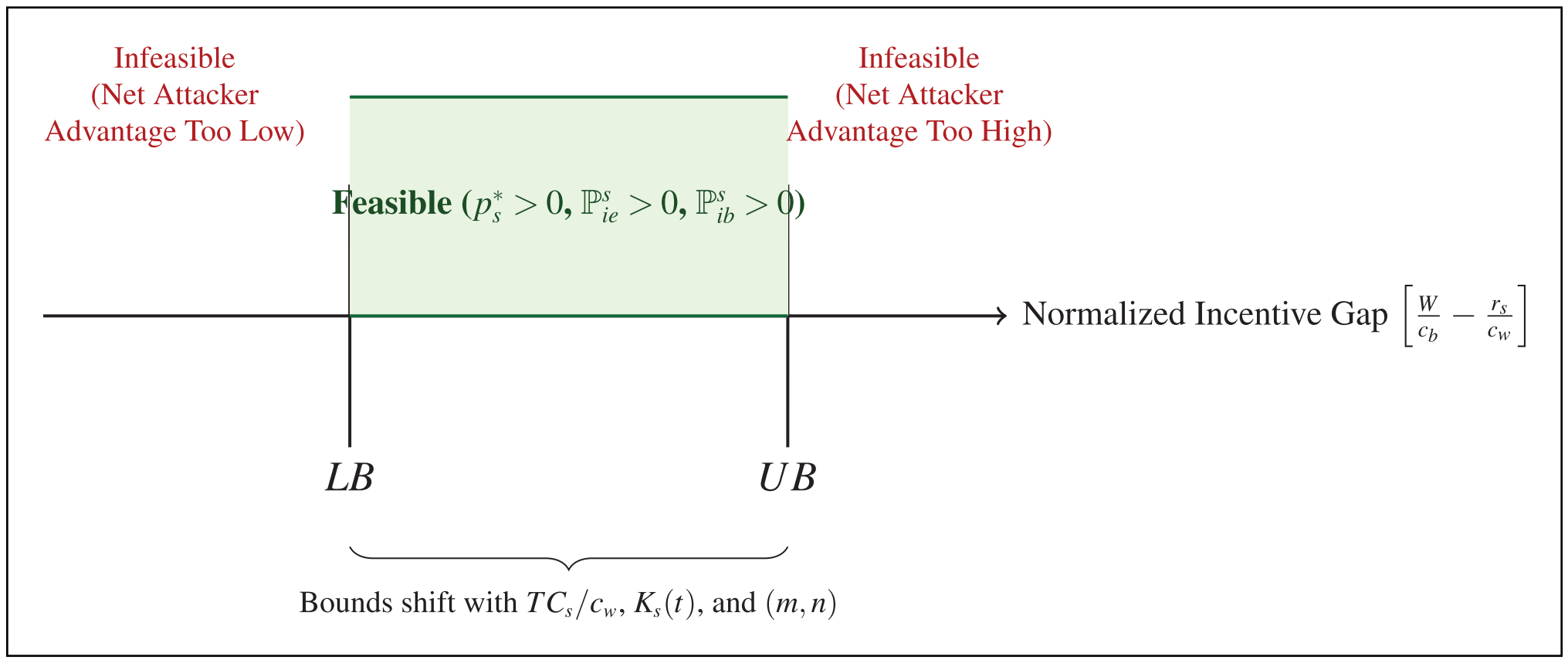

**Figure 3.** Feasibility region for bug bounty programs (BBPs) (Lemma 3). A paid BBP with interior severe-vulnerability discovery outcomes exists if and only if the net normalized incentive gap $\left[\frac{W}{c_b} - \frac{r_s}{c_w}\right]$ lies in $(LB, UB)$ (equation (22)). To the right of *UB*, the net attacker advantage is so large that the bounty required to reallocate first discovery becomes unattractive; to the left of *LB*, the contest does not sustain the interior paid-program equilibrium characterized in the lemma.

Derivations and additional characterization of the bounds are provided in the e-Companion (Section EC.6). In particular, since $UB > LB_1$ and $UB > LB_2$, it follows that $UB > LB = \max\{LB_1, LB_2\}$, so the feasibility set in equation (22) is non-empty. Equations (23) to (25) show that the location of the feasibility interval is shaped by breach losses $TC_s$, release-time-dependent severe exposure $K_s(t)$, and the scale and composition of the severe-bug discovery contest $(m, n)$. Practically, equation (22) serves as a pre-implementation screen: vendors can improve feasibility by lowering normalized attacker gains (e.g., hardening, monitoring, or enforcement that reduces effective illicit gains per unit effort), increasing normalized reputational payoffs (recognition and status mechanisms), and reducing severe exposure at release through targeted pre-release assurance. This feasibility characterization complements prior work on disclosure and security economics by identifying firm-side primitives under which BBPs can reallocate first discovery while sustaining interior outcomes (Cavusoglu et al., 2007; Guda et al., 2021; Kumar and Mallipeddi, 2022; Luo and Choi, 2022).

## 3.10 Release time and firm profit with BBP

To determine the effect of establishing a BBP on the release time of the software and the firm's profits, we start by considering the optimal release time of the software for a vendor without BBP.

*3.10.1 Optimal release time without BBP.* First, note that in the absence of a BBP, $p_s = 0$, and the success probabilities of eWHH and BHH change accordingly as follows:

$$\mathbb{P}_{ie}^s = \max\left(0, \frac{1}{n+m} \times \left[1 + \frac{mK_s(t)}{(n+m-1)(n+m)}\left\{\frac{r_s}{c_w} - \frac{W}{c_b}\right\}\right]\right) \tag{26}$$

$$\mathbb{P}_{ib}^s = \max\left(0, \frac{1}{n+m} \times \left[1 + \frac{nK_s(t)}{(n+m-1)(n+m)}\left\{\frac{W}{c_b} - \frac{r_s}{c_w}\right\}\right]\right) \tag{27}$$

There are potentially three types of costs when SVs are discovered. First, the vendor may incur the cost $TC_{ns}$ if a non-severe SV is discovered by users,[13] which may hurt the vendor's reputation. Second, the vendor incurs the cost $TC_s$ if a BHH discovers and exploits a severe SV. Finally, it may incur the cost $xTC_s$ with $x \in (0, 1)$ when an eWHH discovers a severe SV and potentially discloses it publicly without coordination with the vendor. This uncoordinated disclosure imposes an externality on the software vendor and potentially on the users of its software. Assuming that $x \in (0, 1)$ implies that this cost, $xTC_s$, is not as high as the damage inflicted on the vendor by a BHH who finds the severe SV first. This risk contrasts sharply with the scenario under a BBP. A fundamental function of a BBP is to establish a formal, contractual agreement between the vendor and the WHHs. Participants must adhere to strict rules of engagement and coordinated disclosure policies in exchange for bounties and legal safe harbor. Violating these terms carries legal and significant reputational consequences within the security community. Therefore, we assume that WHHs participating in the BBP comply with controlled disclosure, effectively eliminating the risk captured by $x$ (i.e., $x = 0$ under BBP).

This difference in disclosure control is one of the primary motivations for vendors adopting BBPs.

The objective function of the vendor without a bounty program becomes as follows:

$$\Pi_{nb} = R(t) - K_s(t) m \mathbb{P}^s_{ib} TC_s - K_s(t) n \mathbb{P}^s_{ie} (xTC_s) - \delta K_{ns}(t) TC_{ns} \tag{28}$$

Assuming the interior region where $\mathbb{P}^s_{ib} > 0$ and $\mathbb{P}^s_{ie} > 0$ (so the max operator does not bind), the first-order condition to determine the optimal release time is as follows:

$$\begin{aligned}\frac{\partial \Pi_{nb}}{\partial t} = R'(t) - K_s'(t) \frac{m}{n+m} \\ \times \left[ 1 + \frac{2nK_s(t)}{(n+m-1)(n+m)} \left\{ \frac{W}{c_b} - \frac{r_s}{c_w} \right\} \right] \\ \times TC_s - K_s'(t) \frac{n}{n+m} \\ \times \left[ 1 + \frac{2mK_s(t)}{(n+m-1)(n+m)} \left\{ \frac{r_s}{c_w} - \frac{W}{c_b} \right\} \right] \\ \times (xTC_s) - \delta K_{ns}'(t) TC_{ns} = 0\end{aligned} \tag{29}$$

The assumptions that $R''(t) < 0$ and $K''(t) > 0$ ensure that this first-order condition is also sufficient for maximization, given that the objective is concave in $t$.[14] From the first-order condition as stated in equation (29), we can derive comparative statics on the optimal release time $t$. The optimal release time increases as $TC_s$, $TC_{ns}$, and $x$ increase. Furthermore, the optimal release time increases if the inherent vulnerability risk of the software is higher (i.e., if the function $K_s(t)$ shifts upward, indicating a higher likelihood of residual vulnerabilities for any given testing duration). Since additional testing reduces the likelihood of residual SVs (i.e., $K'(t) < 0$) and thus lowers the expected breach costs, the vendor will delay the release if the cost of severe vulnerability exploitation by BHH is high, if the risk from uncoordinated disclosure by WHH is high, or if the software is inherently more complex or prone to vulnerabilities.

*3.10.2 Profit effects and baseline conditional release-timing implications of BBPs.* Lemma 3 establishes the bounds for $\left[ \frac{W}{c_b} - \frac{r_s}{c_w} \right]$ that are necessary for BBPs to exist, that is, $p_s > 0$, $\mathbb{P}^s_{ie} > 0$, and $\mathbb{P}^s_{ib} > 0$. When the condition in Lemma 3 is met, we can substitute the expressions (20) and (21) obtained for the optimal monetary awards in the objective function (19) of the vendor with BBP. This objective can then be expressed in terms of the objective function (28) of the vendor without BBP, as follows:

$$\begin{aligned}\Pi_b = \Pi_{nb} + \frac{mn\,(K_s(t))^2\, p_s^2}{(n+m-1)(n+m)^2\, c_w} \\ + (K_{ns}(t))^2\, p_{ns}^2 + \; nK_s(t)\, xTC_s \\ \max\left( 0, \; \frac{1}{n+m} \left[ 1 + \frac{mK_s(t)}{(n+m-1)(n+m)} \right.\right. \\ \left.\left. \times \left\{ \frac{r_s}{c_w} - \frac{W}{c_b} \right\} \right] \right)\end{aligned} \tag{30}$$

The following proposition states the effect of the BBP on vendor profits.

**Proposition 2** (BBP Leads to Higher Profits)**.** *Whenever a BBP is feasible (as defined in Lemma 3), adopting the program strictly increases the vendor's expected profit relative to not adopting a BBP.*

Proposition 2 follows from equation (30): Since the second term $mn\,(K_s(t))^2\, p_s^2 / (n+m-1)(n+m)^2\, c_w$ is positive (as $p_s > 0$ from Lemma 3), and the remaining terms are non-negative, we have $\Pi_b > \Pi_{nb}$. Moreover, the program is profitable even if $x = 0$, that is, when there is no risk of uncoordinated disclosure by the eWHH. This highlights the BBP's fundamental mechanism: risk conversion. The program strategically reallocates the first-finder probability on severe vulnerabilities from malicious hackers (BHHs) to ethical hackers (WHHs). In doing so, the vendor converts exposure to catastrophic, unexpectedly large losses ($TC_s$) into controlled, pay-for-results expenditures ($p_s$). While controlled disclosure (reducing $xTC_s$) provides additional value when $x > 0$, the risk conversion mechanism alone drives the profitability increase. This higher profitability helps explain the growing adoption of BBPs among software vendors. The overall cost of bugs first found by BHH ($TC_s$) can be extremely high. This cost includes not only the monetary damage inflicted by hackers but also the regulatory and reputational harm incurred by the vendor. Thus, software vendors would opt to implement BBPs to reduce such costs. This profit-improvement result highlights a distinct mechanism relative to prior models of disclosure and patching: a BBP is a vendor-designed contest that reallocates severe-vulnerability first discovery away from adversaries and converts a portion of breach exposure into bounded, pay-for-results operating expense. Operationally, the mechanism is most valuable when the vendor can sustain timely triage and remediation; otherwise, higher reporting intensity can congest queues without materially reducing exploit risk. We next examine the baseline model's release-timing implication, which we interpret more narrowly in practice.

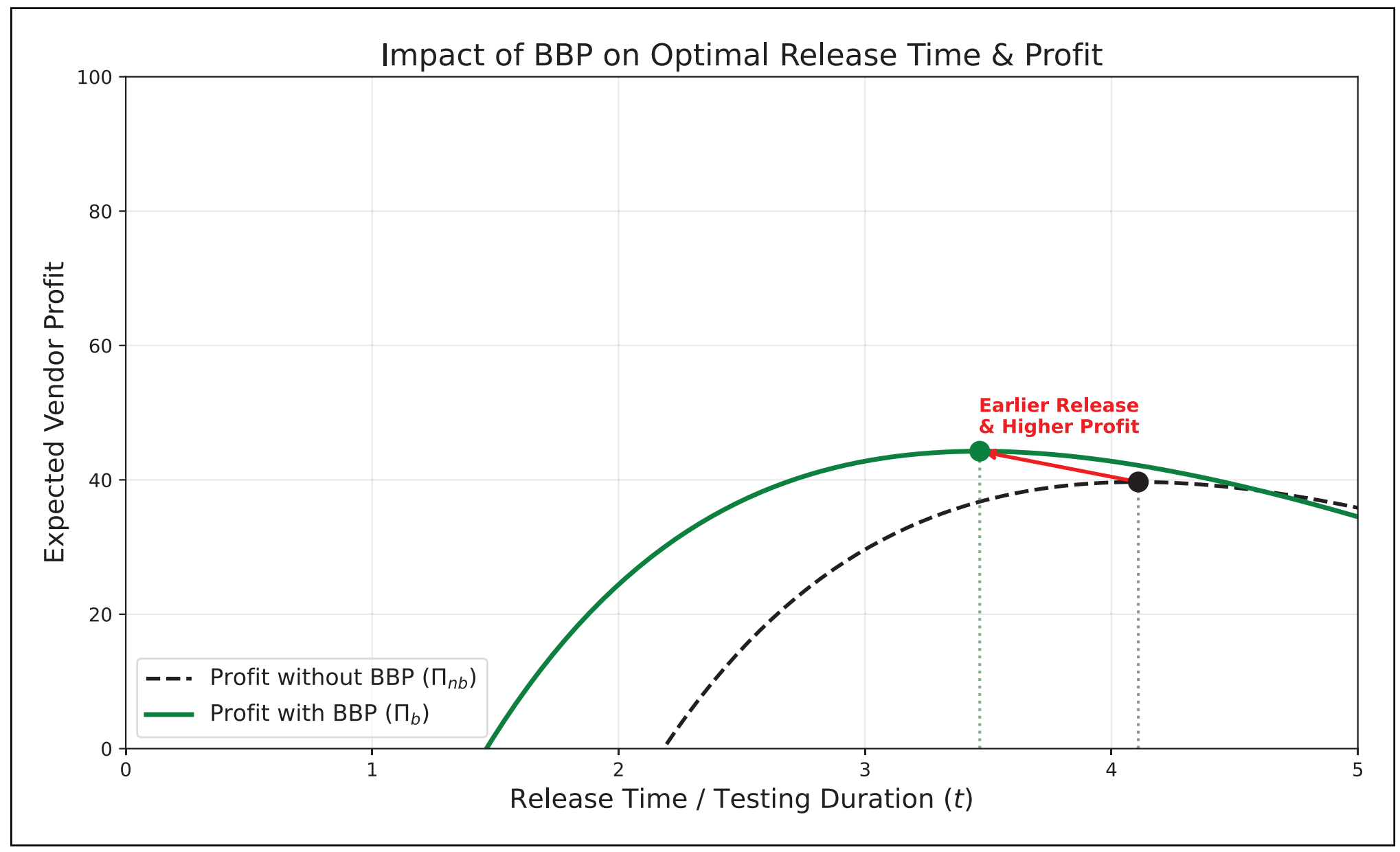

**Figure 4.** Impact of bug bounty program (BBP) on optimal release time and profit in the baseline model. An illustrative numerical visualization consistent with Lemma 3 and Propositions 2 and 3. Within the baseline-model feasibility region, BBP adoption shifts expected profit upward and moves the optimal release time earlier.

**Proposition 3** (Earlier Release With BBP in the Baseline Model)**.** *In the baseline model, whenever a BBP is feasible (as defined in Lemma 3), the vendor optimally releases the software earlier than it would in the absence of a BBP* ($t_b^* < t_{nb}^*$)*.*

Figure 4 provides an illustrative baseline-model visualization of Propositions 2 and 3, showing the within-model result that, inside the feasibility region, BBP adoption shifts the vendor's expected profit curve upward and moves the optimal release time earlier.

The proof, included in e-Companion Section EC.10, establishes a within-model result: in the baseline specification, feasibility of the BBP is conditionally sufficient for earlier release relative to the no-BBP benchmark. The proposition does not formally impose triage congestion, validation delays, or deployment frictions; those practical considerations enter through the discussion below and through e-Companion Section EC.9. We therefore interpret the result narrowly when mapping it to practice. It is not a claim that BBPs broadly replace internal assurance. Rather, the proposition identifies how the BBP's risk-conversion mechanism changes the marginal calculus of delay in the baseline model, whereas the real-world relevance of that margin depends on whether the vendor can validate reports, triage them quickly, and translate valid findings into safely deployable patches without substantial backlog.

By evaluating the vendor's profit maximization condition with BBP at the optimal release time chosen without BBP, we show that the vendor prefers to release earlier when BBP is present. To understand Proposition (3) intuitively, note that the vendor's profit function with a BBP includes additional positive terms arising from optimal bounty rewards that depend on the probability of residual vulnerabilities. These bounty-related terms decrease with delay because, as more testing is carried out, $K_s(t)$ and $K_{ns}(t)$ fall; consequently, the optimal severe-vulnerability bounty $p_s$ declines, and the effective non-severe payout term $K_{ns}(t)p_{ns}$ also declines. Even when $x = 0$, meaning that there is no cost from uncoordinated disclosure by eWHHs, these extra profit components remain present. Mathematically, the derivative of these additional terms with respect to $t$ is negative, so delaying the release diminishes their contribution. The baseline model therefore conditionally predicts earlier release when a BBP is present. Whether this within-model margin is likely to obtain in practice depends on the operational conditions discussed below.

The direction and strength of the timing effect also depend on the shape of launch value and on the program's operating burden. If delay has a steep commercial cost, then governed post-release discovery can make the final units of pre-release delay less attractive. If, however, market conditions or AI-mediated interface bypass flatten the launch-value gradient $R'(t)$, the timing benefit from releasing earlier becomes smaller and may disappear even while the BBP continues to improve profit through first-finder reallocation. Likewise, if low-quality or duplicative submissions raise verification and triage burden, the effective operating cost of the program rises, which weakens both adoption incentives and the appeal of any release acceleration.

**Remark 1** (Remediation Capability and Operational Readiness). E-Companion § EC.9 shows that the profitability and timing implications of a BBP rely on operational readiness. If patching is slow, the vendor suffers prolonged post-discovery exposure, captured there by the term $\omega TC_s$, which attenuates the benefit of converting severe-loss exposure into governed discovery and can overturn the timing margin. AI introduces competing forces on this remediation window. On the one side, AI-assisted engineering can accelerate patch drafting and related diagnosis, which mitigates slow-patching concerns in some environments. On the other side, AI-assisted report generation can increase verification and triage burden, slowing the handoff from report to fix. More fundamentally, even when draft patches are produced quickly, safe deployment in complex or regulated environments typically still requires human review, regression testing, and change-control sign-off, so the effective remediation window compresses but need not vanish. In the limiting case of truly near-instant validation and deployment, the governance value of coordinated disclosure would shrink because the private-information window becomes short; however, the discovery-allocation channel would still remain. On balance, AI-assisted acceleration can attenuate the operational-readiness concern in some instances, but the strategic value of coordinated disclosure remains conditional on the vendor's ability to validate, prioritize, and safely deploy fixes. Consistent with observed practice, pre-release assurance and BBP-driven discovery are complementary in levels. This remark therefore qualifies the practical applicability of Proposition 3: once remediation frictions are made explicit, the within-model conditional earlier-release margin can attenuate or disappear even though the profit and discovery-allocation mechanisms remain.

**Robustness: Fixed operating cost of a BBP.** Our baseline analysis abstracts from fixed operating costs of running a BBP, such as legal safe-harbor design, triage staffing, platform fees, and internal engineering coordination. Introducing a fixed overhead cost $F > 0$ would not change the optimal bounty and release-time comparative statics conditional on adoption, but it would shift the adoption condition to $\Pi_b - \Pi_{nb} \geq F$. Thus, some firms may rationally choose not to adopt a BBP even when the marginal incentive mechanism is effective, because the fixed governance and operating burden outweighs the incremental benefit. If overhead costs increase with participation or report noise, for example, through verification, deduplication, or triage load under AI-assisted submissions, the incentive to restrict entry in private programs would be reinforced. Such frictions can also erode operational readiness by slowing verification and report-to-fix throughput, thereby reducing the practical relevance of the earlier-release implication discussed above and in e-Companion Section EC.9.2.

### *3.11 Optimal scope for private BBPs*

We now consider $n$, the number of eWHHs in a program to be a vendor's choice variable and characterize the optimal number of eWHHs that should participate in a BBP. Such a choice applies primarily to private BBPs. A private BBP allows a vendor to choose the eWHHs that the vendor allows to participate in its BBP. Thus, private BBPs not only allow the vendor to choose the quality of eWHH participants but also the number of such participants. Provided that the objective function is concave in $n$, the solution to the first-order condition yields the optimal value for $n$. Differentiating the objective function by $n$ yields the expression as follows[15]:

$$\frac{\partial \Pi_b}{\partial n} = \frac{mK_s^2(t)(TC_s - p_s)^2}{c_w(n+m)^2}\left[-2n^2 - n(m-1) + m(m-1)\right] \tag{31}$$

Solving the first-order condition, $\frac{\partial \Pi_b}{\partial n} = 0$, for $n$ yields the following solution:

$$n = \frac{\sqrt{9m^2 - 10m + 1}}{4} - \frac{(m-1)}{4} \tag{32}$$

The solution yields three structural insights. First, the optimal invite size $n^*$ scales with the expected threat landscape ($m$). Second, and most notably, the optimal number of invited experts is strictly smaller than the number of adversaries ($n^* < m$). This result arises from the *incentive dilution* effect inherent in winner-take-all contests: expanding the invited pool reduces the individual probability of winning for every white hat, thereby depressing their equilibrium search effort. To maximize aggregate defense, the vendor must limit entry to maintain high individual stakes. Third, as the external threat ($m$) grows, the vendor optimally expands the invited pool ($\partial n^*/\partial m > 0$) to counter the higher malicious discovery rate, but remains "leaner" than the attacker population to preserve effort intensity.

This scope result advances OM research on contest design by providing a security-specific microfoundation for restricted entry. Consistent with innovation tournament theory (see e-Companion Section EC.12.3), restricting the invited pool maintains high individual stakes while simultaneously mitigating operational frictions such as triage congestion and duplicate submissions (Akgul et al., 2023). Managerially, this reframes crowd size as a dynamic operating lever: private BBPs should be sized to the threat ($m$), not to an abstract ideal of "more eyes." High-exposure products warrant larger invite sets, yet the strict $n^* < m$ condition dictates that curation is superior to indiscriminate breadth. Operationally, firms should scale invitations in step with observed threat signals and internal patching capacity; expanding scope without matching remediation throughput risks clogging queues and eroding researcher incentives. Unlike public programs, private BBPs allow vendors to explicitly enforce this efficient scope to maximize protection per bounty dollar.

## 4 Conclusion

This article analyzes how BBPs reshape vendors' security and release choices by embedding coordinated post-release vulnerability discovery and disclosure into the firm's operating problem. In the model, the vendor chooses release time and severity-contingent bounties, while expert and non-expert ethical hackers and malicious attackers compete in winner-take-all discovery races.

The analysis yields four core findings. First, we derive closed-form bounty schedules. The optimal severe-vulnerability bounty increases with the vendor's breach loss and attackers' normalized illicit gains, decreases with ethical hackers' reputational payoffs, and rises with the likelihood of residual severe vulnerabilities at release. The optimal non-severe bounty is proportional to the expected loss from non-severe defects that would otherwise be discovered by users. Second, we characterize a feasibility interval, expressed in normalized illicit and reputational terms, that guarantees positive bounties and interior discovery outcomes. Within this interval, a BBP strictly increases expected vendor profit by reallocating first-discovery probability on severe vulnerabilities away from attackers and toward coordinated ethical reporting, thereby converting catastrophic breach exposure into bounded, pay-for-results expenditures. Third, for private BBPs, the optimal invited set of expert ethical hackers increases with expected threat intensity but remains strictly smaller than the expected number of attackers, reflecting winner-take-all incentive dilution. Fourth, in the baseline model, BBP feasibility conditionally implies earlier release relative to the no-BBP benchmark because governed post-release discovery reduces the marginal benefit of additional delay. This fourth result is intentionally narrow: it is a within-model conditional implication, whereas its practical relevance depends on operational readiness, triage throughput, and safe deployment capacity.

For practice, BBPs are best interpreted as an additional governance layer that complements strong internal assurance while changing how residual vulnerability risk is managed. Managers should calibrate severe bounties against breach-loss models and attacker monetization proxies, actively invest in reputation mechanisms (e.g., public acknowledgment and researcher status) that substitute for cash in attracting high-effort reporting, and, when programs are private, right-size the invited set to preserve effort incentives and limit triage congestion. If scarcity in illicit markets raises the expected monetization of the remaining severe vulnerabilities, the model predicts that vendors must raise severe bounties to remain competitive, although this pressure is naturally capped by the exploitability and economic value of any given bug. Even when the release-timing margin does not carry over to practice, these bounty-design, feasibility, and scope results continue to guide program design. When the baseline model's conditional timing margin is practically relevant, it is most plausible when launch-value losses from delay remain steep, report-processing capacity is not overwhelmed by low-quality or duplicative submissions, and valid reports can be translated into safely deployable fixes without substantial backlog.

These results also have governance and policy implications. Coordinated disclosure internalizes reporting and concentrates vulnerability information with the vendor during remediation, which can reduce exploit risk but also create information asymmetry for users. Governance mechanisms that preserve coordinated reporting while limiting socially costly opacity include: (i) time-bounded disclosure norms and staged disclosure (limited details pre-patch, fuller details post-patch), (ii) standardized severity and reporting formats that reduce classification disputes and improve comparability, and (iii) incentives or requirements tied to remediation service levels and transparent post-patch disclosure. The goal is to retain the efficiency gains from coordinated discovery while limiting user exposure during remediation.

Several limitations suggest directions for future research. The baseline takes hacker populations as exogenous; the e-Companion endogenizes expert participation and expertise selection among ethical hackers (e-Companion Section EC.11), and additional work could jointly endogenize attacker entry and dynamic learning in a unified framework. A second limitation concerns report-processing frictions. Low-quality or duplicative submissions, including AI-assisted report noise, raise verification and triage cost, while AI-assisted engineering can shorten diagnosis and patch drafting. A fuller treatment would model these offsetting forces formally rather than as reduced-form robustness checks. A third limitation concerns the remediation window and vendor reputation. E-Companion Section EC.9 shows that slower patching weakens the benefit of governed discovery; future work could extend this by modeling vendor-side reputational penalties for known-but-unpatched vulnerabilities, particularly when disclosure or leakage occurs before a fix is deployed. Finally, our analysis is firm-centric; a welfare extension would incorporate user losses during nondisclosure and the value of precaution enabled by earlier information, and would study how optimal bounty design interacts with disclosure rules and reporting requirements that internalize these externalities.

### Acknowledgments

The authors gratefully thank the reviewers of POM.

### ORCID iDs

Muhammad Zia Hydari https://orcid.org/0000-0003-4522-326X
Rahul Telang https://orcid.org/0000-0003-1069-1514

### Funding

The authors received no financial support for the research, authorship and/or publication of this article.

## Declaration of conflicting interests

The authors declared no potential conflicts of interest with respect to the research, authorship, and/or publication of this article.

## Supplemental material

Supplemental material for this article is available online (doi: 10.1177/10591478261448668).

## Notes

1. For an example of a disclosure-transparency initiative, Project Zero introduced a 2025 "Reporting Transparency" trial that publicly signals the existence of newly reported vulnerabilities within about one week of vendor notification; see https://perma.cc/8M2L-BVYC (policy description) and https://perma.cc/B29W-GS4F (public tracking page).
2. For brevity, we place several complementary related-literature discussions in the e-Companion (Section EC.12): interdependence, sourcing, and cloud responsibility; BBP evidence in adjacent fields; contest-design analogues; and policy-oriented interventions. Table EC.2 in the e-Companion also summarizes key connections to the POM literature.
3. See https://perma.cc/94W7-YPG3.
4. Severity is continuous in practice (e.g., CVSS). We adopt a binary severe versus non-severe split for tractability and to focus on high-impact vulnerabilities that drive strategic interactions.
5. Although BBPs may be used for pre-beta and beta testing, these are generally specialized and restricted programs.
6. Our reputation measure focuses on the technical prowess of the hackers, which is associated with their ability to discover complex and impactful bugs. BBPs may have a more nuanced definition of reputation. For example, a prominent BBP platform, HackerOne, describes reputation as "your reputation measures how likely your finding is to be immediately relevant and actionable. Reputation is points gained or lost based on report validity. It's weighted based on the size of the bounty and the criticality of the reported vulnerability. Reputation is based exclusively on your track record as a hacker" (see https://perma.cc/6TJ4-EHAT). Despite the emphasis on report validity in HackerOne's reputation definition, both definitions are broadly similar.
7. Severity classification and triage are also imperfect in practice. To capture this in a reduced-form way, suppose a vulnerability that is objectively severe is downgraded (or not verified as severe) with probability $\varphi$; then ethical hackers perceive an expected severe reward of $(1-\varphi)(p_s + r_s)$. This weakens incentives to prioritize severe-vulnerability search and implies that effective programs may require higher posted rewards and/or clearer severity criteria and validation. For analytical clarity, our formal analysis below sets $\varphi = 0$; introducing $\varphi > 0$ simply rescales the effective severe reward and does not change the qualitative comparative statics.
8. Our winner-take-all formulation is most appropriate for *severe* vulnerabilities, where a single first discovery can trigger either exploitation or coordinated remediation. For *non-severe* vulnerabilities, reports are typically more frequent and less strategically contested. Later, in the vendor objective we use $l\mathbb{P}^{ns}_{ine}$ as a reduced-form proxy for the expected (normalized) volume of valid non-severe reports generated by the non-expert pool. Throughout, we therefore interpret $\mathbb{P}^{s}_{ie}$ and $\mathbb{P}^{s}_{ib}$ as first-discovery probabilities, whereas $\mathbb{P}^{ns}_{ine}$ is best read as a per-researcher reporting yield that scales with $K_{ns}(t)$ and $p_{ns}$.
9. While edge cases exist (e.g., collaborative BHHs or low-visibility exploitation by multiple BHHs), the general premise holds.
10. We restrict attention to parameter values for which the implied success probabilities in Lemma 2 are well-defined and lie in the unit interval $[0, 1]$. Equivalently, the effort differentials implied by equilibrium behavior are bounded by aggregate effort so that the reduced-form linear probability mapping is valid. These bounds are consistent with the BBP feasibility conditions characterized later in Lemma 3.
11. The non-severe term in (16) follows from the neWHH equilibrium effort choice in (13) together with the reduced-form mapping between non-severe incentives and reporting yield used in the vendor objective (18).
12. For non-severe vulnerabilities, $l\mathbb{P}^{ns}_{ine}$ should be interpreted as a reduced-form proxy for the expected volume (or normalized rate) of valid non-severe reports generated by the program, which increases with incentives and reduces expected user-discovery losses.
13. We assume that novice users do not have the expertise and incentive to exert the effort necessary to find severe bugs. Non-severe bugs are much easier to find, even by novice users.
14. These conditions are stronger than necessary. In our analysis, we assume concavity holds so that the first-order conditions are sufficient for maximization.
15. Note that the second derivative of the objective with respect to $n$ is negative, guaranteeing that assumed concavity of the objective with respect to $n$. Hence, the first order condition is also sufficient.